\documentclass[conference]{IEEEtran}
\def\isarxiv{1}

\makeatletter
\def\mdseries@tt{m}
\makeatother

\usepackage{amsmath}
\usepackage{booktabs}
\usepackage{bm}
\usepackage{enumitem}
\usepackage{graphicx}
\usepackage{wrapfig}

\usepackage[frozencache=true]{minted}
\usepackage{moresize}
\fvset{linenos, fontsize=\scriptsize, frame=lines, numbersep=4pt, xleftmargin=1em}

\usepackage{etoolbox}
\makeatletter
\AtBeginEnvironment{minted}{\dontdofcolorbox}
\def\dontdofcolorbox{\renewcommand\fcolorbox[4][]{##4}}
\makeatother

\usepackage[hidelinks]{hyperref}  % \url{} and \href{}

\newcommand{\figureref}[1]{\hyperref[fig:#1]{Fig.~\ref*{fig:#1}}}
\newcommand{\tableref}[1]{\hyperref[tab:#1]{Tab.~\ref*{tab:#1}}}
\newcommand{\listingref}[1]{\hyperref[lst:#1]{Lst.~\ref*{lst:#1}}}
\newcommand{\equref}[1]{\hyperref[eq:#1]{Eq.~\ref*{eq:#1}}}
\newcommand{\secref}[1]{\hyperref[sec:#1]{Sec.~\ref*{sec:#1}}}
\newcommand{\coderef}[1]{line~\ref{code:#1}}

\usepackage[per-mode=symbol]{siunitx}
\DeclareSIUnit \bit {bit}
\DeclareSIUnit \bits {bit}
\DeclareSIUnit \byte {Byte}
\DeclareSIUnit \bytes {Bytes}
\DeclareSIUnit \cycle {cycle}
\DeclareSIUnit \cycles {cycles}
\DeclareSIUnit \hz {Hz}
\DeclareSIUnit \op {Op}
\DeclareSIUnit \mac {MAC}
\DeclareSIUnit \operand {operand}
\DeclareSIUnit \operands {operands}
\DeclareSIUnit \transfer {T}
\DeclareSIUnit \cell {cell}

\providecommand{\pgfsyspdfmark}[3]{}
\usepackage{soul}
\usepackage{xcolor}
\usepackage{pdfcomment}
\soulregister\cite7
\soulregister\ref7
\soulregister\pageref7
\soulregister\tableref7
\soulregister\equref7
\soulregister\secref7
\soulregister\figureref7
\definecolor{sfcolor1}{HTML}{1b9e77}
\definecolor{sfcolor2}{HTML}{e7298a}
\colorlet{highlightcolor1}{sfcolor1!30!white}
\colorlet{highlightcolor2}{sfcolor2!20!white}

\usepackage{caption}  % Customization of figure captions
\usepackage{subcaption}
\usepackage{listings}
\usepackage{graphicx}
\DeclareCaptionSubType{listing} % Allow subfigures
\newcommand{\matr}[1]{\bm{#1}}

\begin{document}

\title{Fast Arbitrary Precision Floating Point on FPGA}
% \title{We Dumped an Arbitrary Precision Multiplier\\on an FPGA and It Went Fast}

\author{\IEEEauthorblockN{%
Johannes de~Fine~Licht\IEEEauthorrefmark{1}, %
Christopher A. Pattison\IEEEauthorrefmark{2}, %
Alexandros Nikolaos Ziogas\IEEEauthorrefmark{1},\\ %
David Simmons-Duffin\IEEEauthorrefmark{3}, %
Torsten Hoefler\IEEEauthorrefmark{1}}\vspace{0.5em}
\IEEEauthorblockA{%
\IEEEauthorrefmark{1}Department of Computer Science, ETH Zurich, Switzerland \\
\IEEEauthorrefmark{2}Institute for Quantum Information and Matter, Caltech, Pasadena, USA \\
\IEEEauthorrefmark{3}Walter Burke Institute for Theoretical Physics, Caltech, USA}}

\ifdefined\isarxiv
\else
\IEEEoverridecommandlockouts
\IEEEpubid{\makebox[\columnwidth]{978-1-6654-8332-2/22/\$31.00 \copyright2022 IEEE \hfill}
\hspace{\columnsep}\makebox[\columnwidth]{ }}
\fi

\maketitle

\begin{abstract}
Numerical codes that require arbitrary precision floating point (APFP) numbers for their core computation are dominated by elementary arithmetic operations due to the super-linear complexity of multiplication in the number of mantissa bits.
APFP computations on conventional software-based architectures are made exceedingly expensive by the lack of native hardware support, requiring elementary operations to be emulated using instructions operating on machine-word-sized blocks.
In this work, we show how APFP multiplication on compile-time fixed-precision operands can be implemented as deep FPGA pipelines with a recursively defined Karatsuba decomposition on top of native DSP multiplication. When comparing our design implemented on an Alveo U250 accelerator to a dual-socket 36-core Xeon node running the GNU Multiple Precision Floating-Point Reliable (MPFR) library, we achieve a $\bm{9.8{\times}}$ speedup at $\bm{4.8}$\:GOp/s for 512-bit multiplication, and a $\bm{5.3{\times}}$ speedup at $\bm{1.2}$\:GOp/s for 1024-bit multiplication, corresponding to the throughput of more than $\bm{351{\times}}$ and $\bm{191}{\times}$ CPU cores, respectively.
We apply this architecture to general matrix-matrix multiplication, yielding a $\bm{10{\times}}$ speedup at $\bm{2.0}$\:GOp/s over the Xeon node, equivalent to more than $\bm{375{\times}}$ CPU cores, effectively allowing a single FPGA to replace a small CPU cluster.
Due to the significant dependence of some numerical codes on APFP, such as semidefinite program solvers, we expect these gains to translate into real-world speedups. Our configurable and flexible HLS-based code provides as high-level software interface for plug-and-play acceleration, published as an open source project.
\end{abstract}

\IEEEpeerreviewmaketitle

%%%%%%%%%%%%%%%%%%%%%%%%%%%%%%%%%%%%%%%%%%%%%%%%%%%%%%%%%%%%%%%%%%%%%%%%%%%%%%%%%%%%%%%%%%%%%%%%%%%%%%%%%%%%%%%%%%%%%%%%
\section{Introduction}
%%%%%%%%%%%%%%%%%%%%%%%%%%%%%%%%%%%%%%%%%%%%%%%%%%%%%%%%%%%%%%%%%%%%%%%%%%%%%%%%%%%%%%%%%%%%%%%%%%%%%%%%%%%%%%%%%%%%%%%%

% Introduction to the high-level topic ---------------------------------------------------------------------------------

\noindent Arbitrary precision arithmetic, such as that implemented by the GNU Multiple Precision (GMP)~\cite{gmp} and Multiple Precision Floating-Point Reliable (MPFR)~\cite{mpfr} libraries (where it is referred to as  ``multi-precision'' arithmetic), allows increasing precision by extending the number of bits used to represent numbers beyond the machine word size natively supported by software processors.
This can be necessary to accurately investigate domains where information is found in small differences between numbers (i.e., numbers that are very similar and nearly cancel each other out), which cannot be effectively captured by the dynamic precision of floating-point arithmetic.

% SDP and Conformal bootstrap -----------------------------------------------------------------------------------------
As motivation for this work, we consider semidefinite programs (SDPs). SDPs are ubiquitous and efficiently solvable convex optimization problems involving a linear cost function
of a positive-semidefinite matrix subject to affine constraints \cite{doi:10.1137/1038003}.
SDPs have myriad applications in fields such as control theory, combinatorial optimization, algebraic geometry, and operations research \cite{doi:10.1137/1.9781611970777,doi:10.1137/0805002,HAUENSTEIN2021100166,Wolkowicz2000}. A popular approach to solving the resulting SDPs is primal-dual interior-point methods, which rely on matrix decompositions and matrix-matrix multiplication. However, such methods frequently encounter ill-conditioned matrices, and consequently, several solvers have been implemented to solve SDPs using high-precision arithmetic~\cite{sdpa_gmp, sdpa_campary, cosmo, sdpb}.
A state-of-the-art SDP library is SDPB~\cite{sdpb,Landry:2019qug}, an interior-point solver designed to handle semidefinite programs that arise in the conformal bootstrap. The conformal bootstrap is a powerful framework for studying phase transitions in a wide variety of physical systems~\cite{Simmons-Duffin:2016gjk,Chester:2019wfx,bootstrap_review}. Its central strategy is to solve a series of SDPs to derive rigorous bounds on physical quantities \cite{Rattazzi:2008pe,Poland:2011ey}. In addition to phase transitions, SDPB has been applied to problems in sphere packing \cite{Afkhami-Jeddi:2020ezh}, scattering amplitudes \cite{Paulos:2016but,Caron-Huot:2020cmc}, and differential geometry \cite{Bonifacio:2020xoc,Kravchuk:2021akc}. In all of these cases, the relevant physical or mathematical system satisfies an infinite set of consistency conditions, only a finite subset of which are used in a given SDP. \emph{High-precision arithmetic enables one to easily and robustly obtain stronger constraints} by systematically enlarging the number of consistency conditions (and the size and complexity of the corresponding SDPs).

% Introduction to the technical issue ----------------------------------------------------------------------------------

Unfortunately, moving from $64$-bit machine word arithmetic to arbitrary precision comes at an immense computational cost.
Fundamental operations, such as addition and multiplication, can no longer be implemented with single instructions and must instead be emulated using a (potentially long) sequence of instructions operating on individual machine-word-sized blocks of the number.
As a result, the runtime of numerical codes that require arbitrary precision arithmetic in their core computation can quickly become dominated by elementary arithmetic operations. This is exacerbated by the super-linear complexity of multiplication (and consequently dependent operations such as division) in the number of mantissa bits, which, depending on the instruction mix, can result in arbitrary precision multiplication \emph{alone} dominating workloads such as linear algebra.
While specialized instructions have been introduced to x86 to mitigate this, namely \texttt{ADCX} (add with carry) and \texttt{MULX} (unsigned integer multiplication with double-width output), the issue of emulation and complexity remains.

% The opportunity ------------------------------------------------------------------------------------------------------

The reconfigurable hardware fabric in FPGA devices allows deploying custom circuits in terms of the elementary components available on the chip.
Due to the importance of machine learning workloads, recent work in both fixed and reconfigurable hardware acceleration has focused on low precision types~\cite{finn, bfloat, microsoft_floating_point}. However, FPGAs are also an excellent platform for going in the other direction:
While they often cannot compete with GPUs on accelerating traditional floating-point-dominated workloads, they have a significant advantage on data types that are not natively supported by the instruction sets of other architectures, such as arbitrary precision arithmetic, as these operations can be unrolled and deeply pipelined on the chip.
By accelerating the basic arbitrary precision operators on FPGA, the speedup achieved can then directly translate into real-world speedup in codes that are dominated by arbitrary precision arithmetic.

% the large dependence of the numerical bootstrap and arbitrary precision semidefinite programming on linear algebra routines such as Cholesky decomposition and matrix-matrix multiplication, these performance gains on multiply-accumulate can be translated into real world speedup.
% %
% Using FPGAs, arbitrary precision-dominated workloads like SDPB~\cite{sdpb} could be sped up by 1-2 orders of magnitude.

% Our approach ---------------------------------------------------------------------------------------------------------
% Our contribution -----------------------------------------------------------------------------------------------------

In this work, we show how a Karatsuba-based arbitrary precision floating point (APFP) multiplier implemented on a single FPGA device can outperform $351{\times}$ CPU cores executing MPFR.
We deploy this architecture in a general matrix-matrix multiplication (GEMM) accelerator, which is a crucial component of many numerical workloads, yielding a design that outperforms $375{\times}$ CPU cores.
The accelerator is exposed through a BLAS-like software interface and published as open-source code on GitHub\footnote{\url{https://github.com/spcl/apfp}}, allowing plug-and-play FPGA acceleration of existing APFP-dominated workloads by modifying a few lines of code.
The HLS-based code is highly configurable to support different precisions, tile sizes, FPGA architectures, DRAM layouts and more, and can target any platform supported by Xilinx' Vitis toolflow and the Xilinx Runtime (XRT), including both current and future devices.
Following our approach, this acceleration can be extended to other APFP routines in linear algebra and beyond, providing significant speedup that can enable new science in practice.

%%%%%%%%%%%%%%%%%%%%%%%%%%%%%%%%%%%%%%%%%%%%%%%%%%%%%%%%%%%%%%%%%%%%%%%%%%%%%%%%%%%%%%%%%%%%%%%%%%%%%%%%%%%%%%%%%%%%%%%%
\section{Arbitrary Precision Floating Point Operators}
\label{sec:apfp_ops}
%%%%%%%%%%%%%%%%%%%%%%%%%%%%%%%%%%%%%%%%%%%%%%%%%%%%%%%%%%%%%%%%%%%%%%%%%%%%%%%%%%%%%%%%%%%%%%%%%%%%%%%%%%%%%%%%%%%%%%%%

\noindent The most fundamental arithmetic building blocks of most computations in high-performance computing (HPC) are addition (including subtraction) and multiplication. The most common performance metric used to evaluate and rank HPC systems is their throughput in terms of these two operators. For arbitrary precision-based codes, they are the most critical to accelerate. In the following, we cover our FPGA implementation for APFP addition and multiplication.

We base the functional behavior of our arithmetic on that implemented in the GNU MPFR library, using the round-to-zero mode (\texttt{MPFR\_RNDZ}). In MPFR, an APFP number is implemented as a struct containing four runtime fields: the number of bits used for the mantissa; the sign, stored as a machine word; the exponent, stored as a machine word; and a pointer to a heap-allocated array of ``limbs'', where each limb is a machine word-sized chunk of the mantissa.

\begin{figure}
  \centering
  \begin{minipage}{\columnwidth}
    \centering
    \includegraphics[width=.9\textwidth]{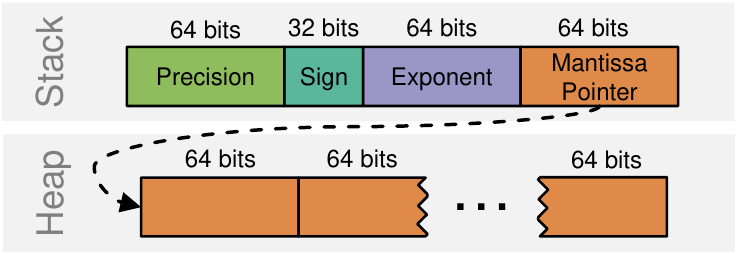}
  \end{minipage}\vspace{1em}
  \begin{minipage}{\columnwidth}
    \centering
    \includegraphics[width=.9\textwidth]{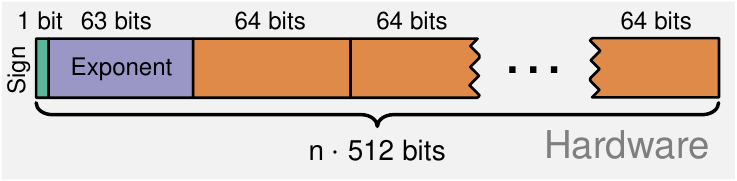}
  \end{minipage}
  \caption{The MPFR software representation on a $64$-bit system (top) transformed into a hardware-friendly format (bottom) for integer $n$, where $n \cdot \SI{512}{\bits} - \SI{64}{\bits}$ represent the mantissa.}
  \label{fig:mpfr_to_apfp}
\end{figure}

To adapt the MPFR representation to a hardware-suitable format, we apply the following changes to the representation, without affecting the functional semantics of the operators:
\begin{itemize}[leftmargin=*, topsep=3pt, itemsep=3pt]
  \item The number of bits used for the mantissa is kept configurable but fixed at compile-time, allowing us to omit this field from the data type at runtime.
  \item The sign is packed into a single bit of the exponent, reducing the exponent to a $(b_\text{limb} - 1)$-bit signed integer (e.g., $63$ bits), where $b_\text{limb}$ is the machine word size that MPFR is configured with (typically $64$ bits).
  \item The mantissa is packed tightly with the sign and exponent rather than being allocated separately, which is possible due to the precision being configured at compile-time.
  \item The combined sign, exponent, and mantissa are packed into a multiple of $512$ bits to enable efficient memory accesses.
\end{itemize}
The format transformation is illustrated in \figureref{mpfr_to_apfp} for a system with $64$-bit machine words.
The provided \texttt{ap\_uint} arbitrary precision integer type in Vitis HLS is used to pack the sign, exponent, and mantissa tightly and ensure that wide buses are generated on the FPGA\@. Our operators will maintain \textbf{full bit-compatibility} in the mantissa with MPFR, and their output will be compared to the equivalent MPFR software computation to verify correctness of the implementation.

%-----------------------------------------------------------------------------------------------------------------------
\subsection{Floating-Point Multiplier}
%-----------------------------------------------------------------------------------------------------------------------

The majority of work involved in floating-point multiplication lies in the underlying unsigned integer multiplication of the two mantissas. Consequently, multiplying the mantissas will account for the majority of hardware utilization in the floating-point multiplication kernel, and the majority of hardware utilization in all the kernels benchmarked in this work.

Naive multiplication of integers (commonly referred to as the ``textbook'' algorithm) requires $O(b^2)$ work in the number of bits $b$ used to represent the integers. However, by recursively decomposing and reorganizing the multiplication into subcomponents, some redundant subcomputations can be eliminated to reduce the asymptotic complexity at the cost of higher constants, first described by Karatsuba~\cite{karatsuba} achieving $O(b^{\text{log}_2 3})$, and later generalized by Toom~\cite{toom} and described by Cook~\cite{cook} (the scheme is now commonly referred to as Toom--Cook multiplication). For very high $b$ (not considered in this work), FFT-based methods become practical~\cite{schonhage-strassen, harvey}.

\begin{figure}
  \centering
  \includegraphics[width=\columnwidth]{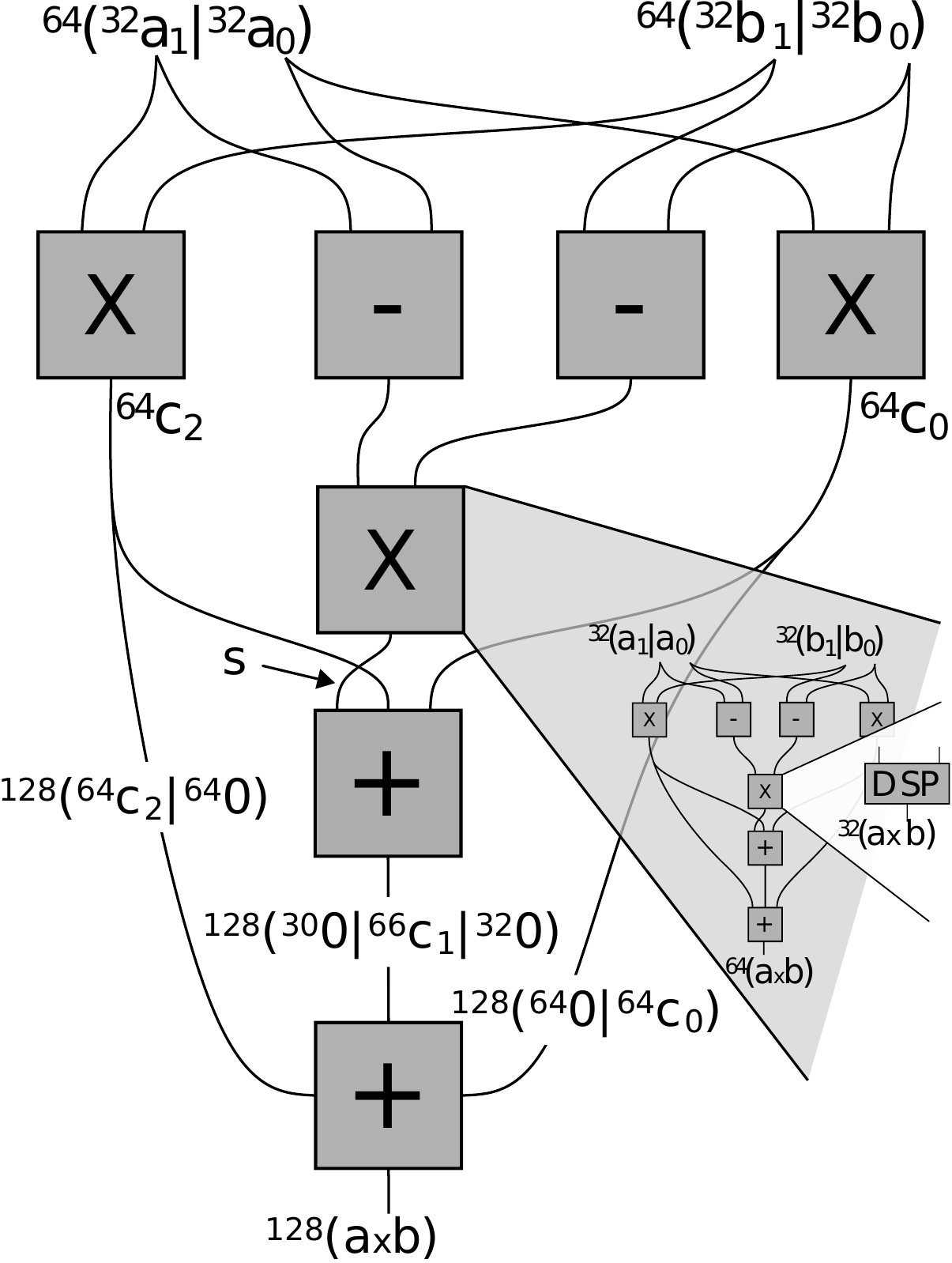}
  \caption{
    Recursive Karatsuba decomposition of $64{\times}64$-bit integer multiplication.
    The explicitly tracked sign bit $s$ in the intermediate computation of $c_1$ is indicated as an arrow.
    Sums with three terms are represented with a single ``$+$'' box.
    Note that the $16{\times}16$-bit multiplication at the lowest level is computed in hardened DSP48E2 units.
  }
  \label{fig:karatsuba}
\end{figure}

In this work, we consider bit widths that are ``large'' from a hardware perspective (i.e., an order of magnitude wider than the $64$-bit words natively supported in CPU and GPU architectures), but ``small'' relative to the overhead imposed by higher-order Toom--Cook and FFT-based methods. To this end, we employ the Karatsuba algorithm for our hardware implementation (also known as Toom--2), which offers a hardware-friendly power-of-two decomposition and is efficient in this middle-ground domain.

The Karatsuba multiplication algorithm and its generalization utilize techniques for fast multiplication of polynomials by viewing the result and operands as a polynomial in a base $B=2^n$ i.e. $a = a_0 + Ba_1$. Here, we describe a single recursive step with bitwidths annoted as superscripts. Each step splits the input operands of $c=ab$ into two halves, the high and low bits ${}^{(2n)}a = {}^{(n)} a_0 + B~{}^{(n)}a_1$, and requires three multiplications at half-bit width per recursive step. Writing $c$ as $c = c_0 + B c_1 + B^2 c_2$ and $b$ as ${}^{(2n)}b = {}^{(n)} b_0 + B~{}^{(n)}b_1$, the coefficients are computed as follows:
\[{}^{(2n)}c_0 = {}^{(n)}a_0 {}^{(n)}b_0\]
\[{}^{(2n)}c_2 = {}^{(n)}a_1 {}^{(n)}b_1\]
\[{}^{(2n)}t =  {}^{(n)}\lvert a_1 - a_0\rvert {}^{(n)} \lvert b_1 - b_0\rvert \]
\[s = \mathrm{sign}\left[(a_1 - a_0)(b_1 - b_0)\right]\]
\[{}^{(2n+2)}c_1 = {}^{(2n)}c_0 + {}^{(2n)}c_2 - {}^{(1)}s~{}^{(2n)}t\]
Explicitly tracking the sign bit in the computation of $c_1$ allows all multiplications to be carried out at $n$ bits. Note that only one multiplication per coefficient is required ($c_0$, $c_2$, and $t$). We recombine the outputs using multiplication by $B$ implemented as shifts. Since a product cannot have more than double the number of digits of the operands, one can see that this addition will not overflow. Combining the contributions yields:
\[{}^{(4n)}c = {}^{(2n)}c_0 + B~{}^{(2n+2)}c_1 + B^2~{}^{(2n)}c_2\]
The decomposition may then be repeated iteratively for the three half-bit width multiplications until reaching a small enough bit width to perform the multiplication as a primitive operation. Our recursive implementation of the decomposition is sketched for $64{\times}64$-bit example inputs in \figureref{karatsuba}.

DSPs in modern FPGAs can natively (and thus efficiently) perform integer multiplication up to a given bit width, and will be used when ``bottoming out'' our decomposition. The DSP48E2 units on the Xilinx UltraScale+ architecture support $18{\times}18$-bit integer multiplication. On this architecture, we thus recursively split the domain until the subcomponents are at most ${\leq}18$~bits in size, after which they can be directly dispatched to DSP units rather than being decomposed further. However, the bottom out bit width is left as a configuration parameter, as falling back on $O(n^2)$ multiplication at a higher bit width can be beneficial (see \secref{config_tuning}).

To implement the Karatsuba decomposition in a general manner to support any input bit width, we exploit C++ template metaprogramming to define a static template recursion that bottoms out on bit widths under the defined bottom out width using an SFINAE~\cite{cpp_templates} pattern. This is illustrated in \listingref{static_recursion}, where the \texttt{ap\_uint} type is used to represent arbitrary bit widths, and \texttt{MULT\_BASE\_BITS} is the chosen threshold where Karatsuba falls back on naive multiplication using DSPs (we will optimize the choice of this threshold in \secref{config_tuning}).

\begin{listing}
  \begin{minted}{C++}
template <int bits>
auto Karatsuba(ap_uint<bits> const &a,
               ap_uint<bits> const &b) ->
    typename std::enable_if<(bits > MULT_BASE_BITS),
                            ap_uint<2*bits>>::type {
  using Full = ap_uint<bits / 2>;
  using Half = ap_uint<bits / 2>;
  Half a0 = a(bits/2-1, 0); Half a1 = a(bits-1, bits/2);
  Half b0 = b(bits/2-1, 0); Half b1 = b(bits-1, bits/2);
  Full c0 = Karatsuba<bits / 2>(a0, b0); // Recurse
  Full c2 = Karatsuba<bits / 2>(a1, b1); // Recurse
  // ...compute |a1-a0| and |b1-b0|...
  Full c1 = Karatsuba<bits / 2>(a1_a0, b1_b0); // Recurse
  // ...combine all contributions and return...
}

template <int bits>
auto Karatsuba(ap_uint<bits> const &a,
               ap_uint<bits> const &b) ->
    typename std::enable_if<(bits <= MULT_BASE_BITS),
                            ap_uint<2*bits>>::type {
  return a * b; // Bottom out using naive mult
}
  \end{minted}
  \caption{Static recursion pattern implemented in C++ bottoming out at \texttt{MULT\_BASE\_BITS} with SFINAE used to implement Karatsuba decomposition for arbitrary bit widths.}
  \label{lst:static_recursion}
\end{listing}

When combining contributions to the final mantissa, we perform integer additions on bit widths up to  $2{\times}$ the number of input bits (e.g., $1024$-bit operands for $512$-bit numbers). Vitis~HLS~2021.2 allows splitting the adder into multiple stages using the \texttt{BIND\_OP} pragma, but only allows a maximum of 4 additional pipeline stages. To avoid deep combinatorial logic and aid routing, we implement an additional pipelined addition/subtraction function that partitions the wide additions into chunks of a configurable base width. We use this to make sure that no more than a fixed number of bits are added in a single cycle, and will show how this impacts resource usage and frequency in \secref{config_tuning}.

%-----------------------------------------------------------------------------------------------------------------------
\subsection{Floating-Point Adder}
%-----------------------------------------------------------------------------------------------------------------------

\noindent Addition of mantissas can be accomplished in a time complexity linear in the number of bits. In the same way as for adding up contributions in Karatsuba multiplication, we partition the integer addition into a configurable number of stages.
To perform a \emph{floating-point} addition, we shift the operands by the difference of the exponents before passing them into the integer adder. Due to the sign-magnitude format of the floating-point format, we must explicitly subtract the operands when the signs differ. When subtracting, the output may become denormalized, requiring us to left-shift the resulting mantissa such that the most significant bit of the mantissa is set, requiring us to count the number of introduced leading zeros and dynamically shift by this number.
% Notably, MPFR differs from the IEEE~754 floating-point format by explicitly storing the leading mantissa bit, which we mimic in our hardware implementation.

We combine the floating-point adder with the multiplier to form a combined multiply-addition pipeline, which can serve as a building block for dense linear algebra kernels.

%-----------------------------------------------------------------------------------------------------------------------
\section{Arbitrary Precision Matrix Multiplication}
\label{sec:gemm}
%-----------------------------------------------------------------------------------------------------------------------

With a fully pipelined multiply-addition unit that performs one operation per cycle, DRAM bandwidth is no longer sufficient to saturate the compute in a linear streaming computation. We thus need to increase the granularity of acceleration to routines that enable sufficient reuse to keep the compute saturated from buffers in on-chip memory. For use in the SDP solvers that motivate this work, general matrix multiplication (GEMM) and derived routines such as the symmetric rank-k update (SYRK) BLAS routine are major workhorses that can provide the necessary reuse.

We design a GEMM architecture that implements the operation $\matr{C} = \alpha \matr{A} \matr{B} + \beta \matr{C}$, where $\matr{A}$ is an $N{\times}K$ matrix, and $\matr{B}$ is a $K{\times}M$ matrix. For the purpose of this work, we fix $\alpha = \beta = 1$, but other values can be introduced at the cost of requiring additional multiplication pipelines, which would correspond to a nearly full replication of the circuit. Reuse is achieved through a 2D tiling scheme, where columns of size $T_N$ from $\matr{A}$ and rows of size $T_M$ from $\matr{B}$ are loaded and used to compute a $T_N {\times} T_M$ outer product, which is accumulated into an output tile of size $T_N \cdot T_M$ of matrix $\matr{C}$ stored in on-chip memory. This is repeated for the full common matrix dimension $K$ until the output tile is complete and is written back to off-chip memory. By setting $T_N = T_M$ and maximizing this quantity, we can achieve optimal fast memory usage in terms of the on-chip memory used~\cite{gemm_fpga}, with an arithmetic intensity of $\frac{T_N T_M}{T_N + T_M}$ ($T_N T_M$ computations for each $T_N$ + $T_M$ operands loaded from memory).

With the outer product scheme selected, one of the input matrices will be read column-wise, while the other will be read row-wise. For the matrix that is not read contiguously from memory, the accesses to DDR memory are less efficient as a result. Fortunately, because each entry occupies a much larger space in memory than traditional data types, even this suboptimal access pattern results in burst reads at least as wide as the floating point number. This is chosen as a multiple of $512$ bits to match the $4{\times}$ clock multiplier of DDR4 memory, the $2{\times}$ data rate, and the $64$-bit DDR4 interface.

When permitted by available resources and routing constraints, we can instantiate multiple GEMM compute units to improve overall throughput. Each compute unit will operate on a distinct partition of the output matrix, such that multiple GEMM accelerators collaborate on a single virtual GEMM call. For $P$ compute units, $N/P$ rows of the input matrix $A$ and the output matrix $C$ are allocated per accelerator and copied to the respective DRAM bank, while the full $B$-matrix is used by every compute unit to compute a complete set of $N/P$ rows of the output matrix.

%%%%%%%%%%%%%%%%%%%%%%%%%%%%%%%%%%%%%%%%%%%%%%%%%%%%%%%%%%%%%%%%%%%%%%%%%%%%%%%%%%%%%%%%%%%%%%%%%%%%%%%%%%%%%%%%%%%%%%%%
\section{Artifacts and Workflow}
%%%%%%%%%%%%%%%%%%%%%%%%%%%%%%%%%%%%%%%%%%%%%%%%%%%%%%%%%%%%%%%%%%%%%%%%%%%%%%%%%%%%%%%%%%%%%%%%%%%%%%%%%%%%%%%%%%%%%%%%

We publish our HLS-based accelerator and the software integration code as open source software, to facilitate it being exploited in APFP-based numerical codes. The hardware accelerator is highly configurable, and once the appropriate bitstream has been built and installed, can be accessed through a high-level BLAS interface, or through CUDA-like device interaction for more fine-grained control.

\subsection{Hardware Accelerator Configuration}

Both software and hardware of our project is configured via CMake. Dependencies required to build the code are automatically detected, including the Xilinx toolchain as enabled by \texttt{FindVitis.cmake} provided by the hlslib~\cite{hlslib} project, which also provides build targets for hardware and hardware emulation for our kernels.

As of writing, the matrix multiplication accelerator can be configured with the following parameters that customize its resource utilization and performance characteristics:
\begin{itemize}[leftmargin=*, topsep=3pt, itemsep=3pt]
  \item \texttt{APFP\_BITS} configures the number of bits used to represent floating point numbers, which includes the bits spent on exponent and sign (packed according to \figureref{mpfr_to_apfp}).
  \item \texttt{APFP\_COMPUTE\_UNITS} sets the replication factor of the multiply-addition pipeline, allowing performance to be scaled up with available resources on the target device.
  \item \texttt{APFP\_TILE\_SIZE\_N} and \texttt{APFP\_TILE\_SIZE\_M} configure the rows and columns of the output tile \emph{per instantiated compute unit}, respectively, increasing memory reuse/reducing memory bandwidth at the cost of on-chip memory resources, as described in \secref{gemm}.
  \item \texttt{APFP\_MULT\_BASE\_BITS} and \texttt{APFP\_ADD\_BASE\_BITS} configure the bit width at which the Karatsuba decomposition falls back on naive multiplication of operands, and the number of bits added in combinatorial logic per pipeline stage when performing wide additions, respectively.
\end{itemize}
By adapting these configuration options to the specific architecture being targeted, the accelerator can be tailored to fully exploit available resources, including logic resources and DRAM banks, and best utilize the characteristics of the underlying hardware components.

\subsection{System Integration}

To make it easy for numerical codes to exploit FPGA acceleration, we expose our accelerator as a high-level software library with BLAS-like API calls. The BLAS interface permits the FPGA acceleration to be a drop-in replacement for libraries such as MLAPACK~\cite{mlapack} or Elemental~\cite{elemental} when the transfer overhead is small relative to the computation size.

\begin{listing}
  \begin{minted}[escapeinside=||]{C++}
El::DistMatrix<El::BigFloat> distr_a = ...;
El::DistMatrix<El::BigFloat> distr_b = ...;
El::DistMatrix<El::BigFloat> distr_c = ...;

// Elemental GEMM
El::Gemm(El::NORMAL, El::NORMAL, El::BigFloat(1), |\label{code:elemental_gemm}|
    distr_a, distr_b, El::BigFloat(1), distr_c);

// Obtain local copies
using LocalMatrix = |\label{code:local_elemental_matrix}|
    El::DistMatrix<El::BigFloat, El::CIRC, El::CIRC>;
LocalMatrix local_a = distr_a;
LocalMatrix local_b = distr_b;
LocalMatrix local_c = distr_c;

// Indexing functions into the matrices
using CIdxF = std::function<mpfr_srcptr(unsigned long)>; |\label{code:indexing_functions}|
using IdxF = std::function<mpfr_ptr(unsigned long)>;

CIdxF index_A = [&](unsigned long i) {
  return local_a.Matrix().Buffer()[i].LockedPointer();
};

// ...define index_B and index_C...

// APFP Interface GEMM Call
apfp::Gemm(apfp::BlasTrans::normal, |\label{code:apfp_gemm}|
    apfp::BlasTrans::normal, m, n, k,
    index_A, local_a.Matrix().LDim(),
    index_B, local_b.Matrix().LDim(),
    index_C, local_c.Matrix().LDim()));
  \end{minted}
  \caption{Example \texttt{GEMM} call for Elemental and for the BLAS compatibility interface. CPU codes relying on Elemental can be converted piece-by-piece by retaining the Elemental data structures.}
  \label{lst:blas_call}
\end{listing}

Elemental is a distributed memory dense linear algebra library, which supports arbitrary precision data types relying on MPFR data types, and uses MPI for parallelization and multi-node support. Using our BLAS interface, we are able to non-invasively accelerate a \texttt{GEMM} call in an Elemental program with minimal additional code (\listingref{blas_call}). The BLAS interface accepts a pointer to a buffer or an \texttt{std::function}/lambda function, accepting an integer and returning an MPFR pointer. This flexibility permits us to avoid copying MPFR data out of the Elemental datatypes while simultaneously avoiding a leaky abstraction with respect to our internal packed floating-point format. The MPFR datatype stores limbs on the heap, so the extra indirection imposed by the indexing function is not a significant drawback.

In \listingref{blas_call}, we show a standard GEMM call in Elemental (\coderef{elemental_gemm}) operating on distributed matrices in addition to a call to the FPGA BLAS interface (\coderef{apfp_gemm}). In this example, the operands are distributed matrices, so they are copied to a single node using the \texttt{El::CIRC} distributed matrix distribution argument (\coderef{local_elemental_matrix}). The only additional code is to define indexing functions that abstract away the layout of the underlying MPFR numbers inside of Elemental (\coderef{indexing_functions}).

While this example copies a distributed matrix to a single MPI process, the Elemental library could be used to facilitate a distributed, multi-FPGA computation.

When data movement to/from the accelerate must be explicitly managed, we provide a fine-grained interface exposing a CUDA-like API to launch kernels and move data between host and device. Workloads with many small matrices will need to keep operands on the FPGA for multiple kernel invocations to amortize the transfer time.

%%%%%%%%%%%%%%%%%%%%%%%%%%%%%%%%%%%%%%%%%%%%%%%%%%%%%%%%%%%%%%%%%%%%%%%%%%%%%%%%%%%%%%%%%%%%%%%%%%%%%%%%%%%%%%%%%%%%%%%%
\section{Evaluation}
%%%%%%%%%%%%%%%%%%%%%%%%%%%%%%%%%%%%%%%%%%%%%%%%%%%%%%%%%%%%%%%%%%%%%%%%%%%%%%%%%%%%%%%%%%%%%%%%%%%%%%%%%%%%%%%%%%%%%%%%

We evaluate our architecture on a Xilinx Alveo U250 accelerator, where we utilize 1--4 DDR4 DRAM banks with a peak bandwidth of $\SI{19.2}{\giga\byte\per\second}$ per bank. The C++-based kernels are implemented in Vitis~HLS with hlslib~\cite{hlslib} extensions, and compiled for hardware with Vitis/Vivado~2021.2, targeting the \texttt{xilinx\_u250\_gen3x16\_xdma\_3\_1\_202020} shell through the OpenCL-based interface relying on the Xilinx Runtime (XRT) version 2.9.317 for host/device interaction. The U250 consists of $4{\times}$ chiplets called ``Super Logical Regions'' (SLRs) that have limited connectivity between them. We thus force kernel instantiations to stay within the bounds of a chiplet to avoid frequency degradation.

To compare against software, we run APFP computations in software using MPFR~4.1.0 and GMP~6.2.1. For dense linear algebra, we run commit~6eb15a0 of Elemental\footnote{\url{https://github.com/elemental/Elemental}}~\cite{elemental} with MPFR/GMP and MPI support.
Benchmarks are run on Cray XC40 compute nodes on the Piz~Daint supercomputer at the Swiss National Supercomputing Center (CSCS),
where each node is equipped with $2{\times}$ Intel Xeon E5-2695 v4 18-core CPUs in a dual-socket configuration (36 cores per node).
The Broadwell-based CPU supports the specialized \texttt{ADCX} add-with-carry instruction from the Intel~ADX x86 instruction set extension targeting arbitrary-precision arithmetic, as well as \texttt{MULX} instruction from the BMI2 extension for $64{\times}64$-bit multiplication with $128$-bit output.
GMP, MPFR, and Elemental are compiled directly on the compute nodes with (Cray) GCC 10.3.0 to exploit these and other architecture-specific optimizations.
Elemental is built with Cray-MPICH 7.7.16. MPI processes are fixed to CPU cores through Slurm to avoid rescheduling of threads across the NUMA boundary.

%-----------------------------------------------------------------------------------------------------------------------
\subsection{Tuning the Multiplier for Resources and Frequency}
\label{sec:config_tuning}
%-----------------------------------------------------------------------------------------------------------------------

\noindent When configuring the APFP multiplier, there are two tunable parameters that represent a trade-off between frequency and resource usage: the threshold at which the Karatsuba decomposition bottoms out and calls naive multiplication using DSPs (\texttt{APFP\_MULT\_BASE\_BITS}); and the number of bits added/subtracted in a single pipeline stage when adding up contributions (\texttt{APFP\_ADD\_BASE\_BITS}). To find the best configurations, we perform a full sweep of this design space for a single 512-bit APFP multiplier, and use this to guide our other experiments. We choose the number of configurable logic blocks (CLBs) as the metric for resource usage, as this is the most utilized resource in our designs, and captures both LUT and register usage. This results in a 2D design space (multiplication and addition configuration) with two evaluation metrics (frequency and CLBs used).

\begin{figure}
  \includegraphics[width=\columnwidth]{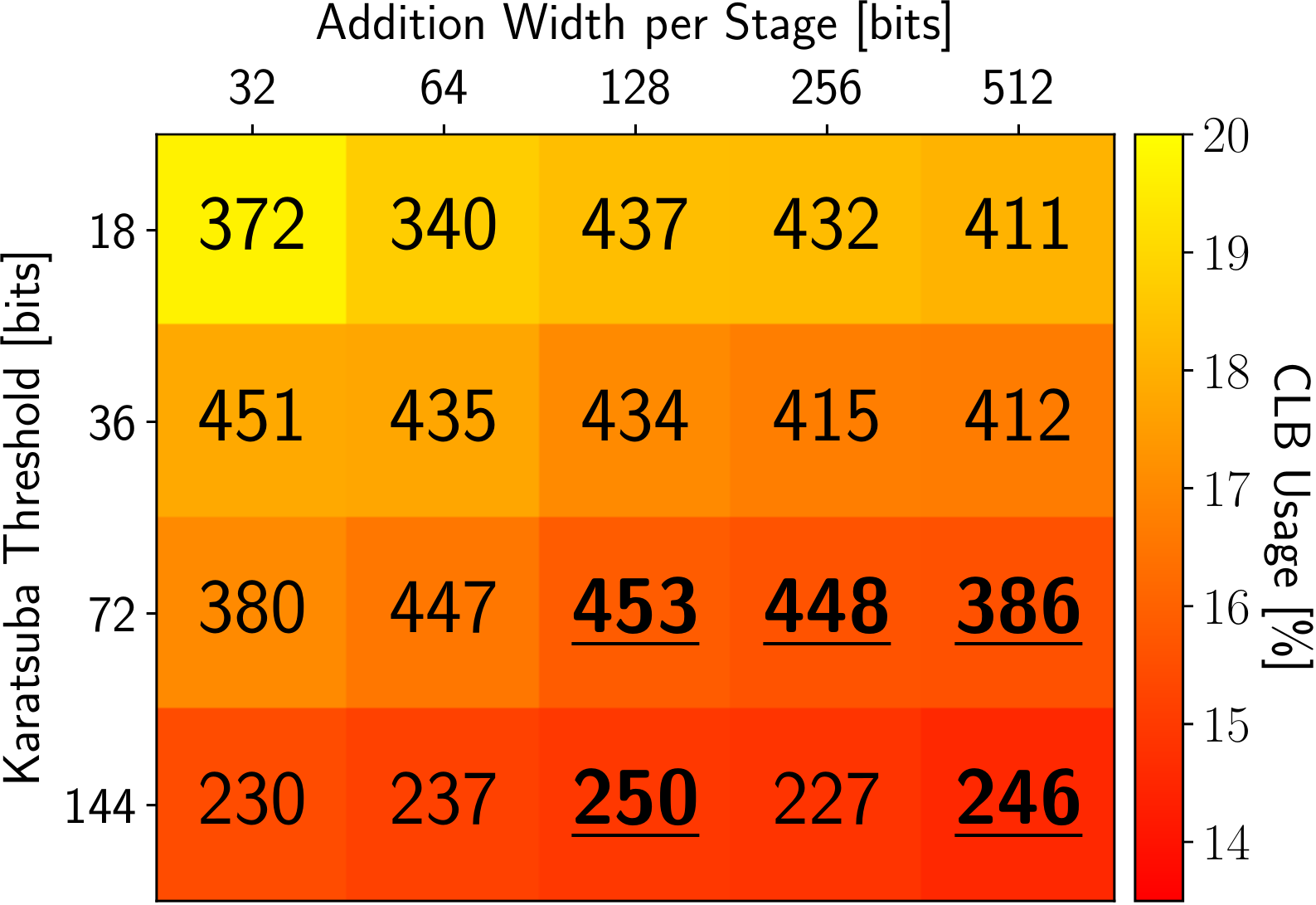}
  \caption{Resource utilization (on the color scale, where brighter colors use more resources) and frequency (annotated on each rectangle, in $\si{\mega\hertz}$) for different number of bits added per pipeline stage, and different thresholds for falling back from Karatsuba onto DSP-based naive multiplication. Pareto efficient configurations are marked in \underline{\textbf{underlined bold font}}.}
  \label{fig:heatmap}
\end{figure}

\figureref{heatmap} shows resource utilization (on the color scale) and frequency (annotated) for different combinations of addition and multiplication configurations for the Karatsuba-based multiplier. For multiplication, the best results are obtained when falling back on DSP-based naive multiplication after 72 bits (lowest resource usage with high frequencies), or 36~bits (consistently high frequencies, but higher resource usage). At 144 bits, the naive multiplication significantly hampers the achievable frequency, while 288~bits fails synthesis altogether. For addition, the best results are obtained when bottoming out at more than 64 bits per pipeline stage. We will target permutations of these configurations of widths for obtaining the best results in the experiments below.

% \begin{table}
%   \centering
%   \begin{tabular}{r | r r r}
%     \textbf{bits/cycle} & \textbf{Frequency} & \textbf{LUTs} & \textbf{Registers} \\ \toprule
%       $32$ bits & $\SI{372}{\mega\hertz}$ & $7.8\%$ & $8.7\%$ \\
%       $64$ bits & $\SI{340}{\mega\hertz}$ & $7.6\%$ & $8.3\%$ \\
%       $128$ bits & $\SI{437}{\mega\hertz}$ & $7.5\%$ & $7.9\%$ \\
%       $256$ bits & $\SI{432}{\mega\hertz}$ & $7.6\%$ & $7.7\%$ \\ \bottomrule
%   \end{tabular}
%   \caption{Frequency and resource usage for different configurations of pipelined addition.}
%   \label{tab:pipelined_add}
% \end{table}

%-----------------------------------------------------------------------------------------------------------------------
\subsection{Benchmarking Floating-Point Multiplication}
%-----------------------------------------------------------------------------------------------------------------------

\noindent To evaluate and compare the performance of the APFP multiplication in isolation, we construct a microbenchmark for both FPGA and CPU that streams from two arrays of operands through the multiplier and writes to an output array in a purely linear fashion.
In this setting, a fully pipelined FPGA multiplier will quickly become memory bound, as it requires $2$ reads and $1$ write per cycle, which corresponds to $\SI{57.6}{\giga\byte\per\second}$ for a \emph{single} 512-bit pipeline at $\SI{300}{\mega\hertz}$, or $\SI{115.2}{\giga\byte\per\second}$ for a single 1024-bit pipeline. Two compute units would thus already grossly exceed the $\SI{76.8}{\giga\byte\per\second}$ peak memory bandwidth of the U250. To evaluate the performance when the compute \emph{can} be fully saturated through memory reuse and/or higher memory bandwidth, we artificially removed the memory bottleneck for the sake of this comparison, by repeatedly feeding the same single data element to the computational kernel. Similarly, although we expect the CPU to primarily be compute bound when running MPFR, we negate any impact from cache misses by constructing the benchmark such that it loops over a dataset that fits in the L1 cache of each Xeon core to ensure the highest possible multiplication throughput for our comparison, representing its true peak running MPFR.

\begin{figure}
  \centering
  \includegraphics[width=\columnwidth]{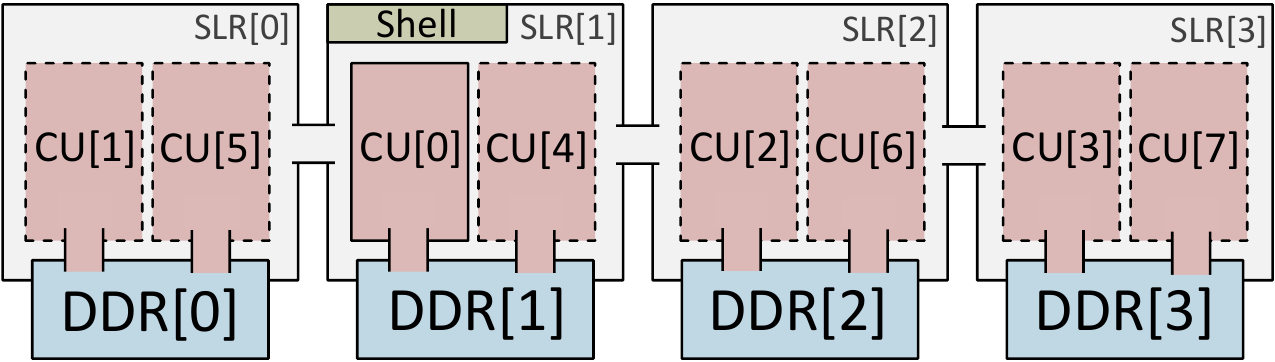}
  \caption{Example mapping of compute units to SLRs/DDR banks on the U250. Only \texttt{CU[0]} is functionally required (solid outline). Round robin continues after the first 8 CUs.}
  \label{fig:compute_unit_mapping}
\end{figure}

For the FPGA accelerator, we replicate the multiplication pipeline to increase the utilization of the FPGA and partition the input problem across the replications. Each compute unit is assigned to a DDR bank in a round-robin fashion, resulting in each unit being assigned to a distinct SLR (chiplet) on the device. We start at DDR bank 1 where the logic interacting with the host is located, then cycle through 0, 2, and 3. Once a compute unit has been assigned to each bank/SLR, the assignment repeats from the first bank. The SLR/bank assignment is illustrated in \figureref{compute_unit_mapping} for up to 8 compute units.

%We test MPFR performance on the CPU in two different configurations. The first one (compute) measures the compute-bound by repeatedly multiplying the same operands. The second configuration (stream) loads, for each multiplication, different operands from large arrays that do not fit in the cache. Furthermore, we run two versions for each of the above configurations, one with simple multiplication and another utilizing fused multiply-add (FMA) operations.

% \begin{figure}
%   \includegraphics[width=\columnwidth]{apfp-artifacts/plots/micro_perf.448.pdf}
%   \caption{Microbenchmark performance  with $448$-bit mantissas ($512$ bits total).}
%   \label{fig:micro512}
% \end{figure}

\begin{table}[b]
  \centering
  \resizebox{\columnwidth}{!}{\begin{tabular}{l | r r r r r r}
  \textbf{Configuration} & \textbf{Freq.}          & \textbf{CLBs} & \textbf{DSPs} & \textbf{Throughput} & \textbf{Speedup} & \textbf{\#Cores} \\ \toprule
  36-core CPU            & $\SI{2100}{\mega\hertz}$ & -        & -        & $\SI{490}{\mega\op\per\second}$  & $1.0{\times}$   & $36{\times}$     \\ \midrule
  FPGA 1 CU              & $\SI{456}{\mega\hertz}$ & $16\%$ & $4\%$  & $\SI{451}{\mega\op\per\second}$  & $0.9{\times}$ & $33.1{\times}$   \\
  FPGA 4 CUs             & $\SI{376}{\mega\hertz}$ & $37\%$ & $14\%$  & $\SI{1502}{\mega\op\per\second}$ & $3.1{\times}$ & $110.3{\times}$   \\
  FPGA 8 CUs             & $\SI{300}{\mega\hertz}$ & $48\%$ & $28\%$ & $\SI{2401}{\mega\op\per\second}$ & $4.9{\times}$ & $176.3{\times}$  \\
  FPGA 12 CUs            & $\SI{300}{\mega\hertz}$ & $62\%$ & $42\%$ & $\SI{3595}{\mega\op\per\second}$ & $7.3{\times}$ & $264.0{\times}$  \\
  FPGA 16 CUs            & $\SI{300}{\mega\hertz}$ & $75\%$ & $56\%$ & $\SI{4784}{\mega\op\per\second}$ & $9.8{\times}$ & $351.3{\times}$  \\ \bottomrule
  \end{tabular}}
  \caption{Our 512-bit (448-bit mantissa) floating-point multiplier executed in hardware, compared to MPFR executed fully in L1 cache on a 36-core CPU node. \#Cores denotes speedup over a single core (i.e., equivalent number of CPU cores).}
  \label{tab:microbenchmark512}
\end{table}

\begin{table}
  \centering
  \resizebox{\columnwidth}{!}{\begin{tabular}{l | r  r r r r r}
  \textbf{Configuration} & \textbf{Freq.} & \textbf{CLBs} & \textbf{DSPs} & \textbf{Throughput} & \textbf{Speedup} & \textbf{\#Cores} \\ \toprule
  36-core CPU & -                       & -        & -        & $\SI{227}{\mega\op\per\second}$  & $1{\times}$   & $36{\times}$     \\ \midrule
  FPGA 1 CU   & $\SI{361}{\mega\hertz}$ & $27\%$ & $8\%$  & $\SI{361}{\mega\op\per\second}$  & $1.6{\times}$ & $57.3{\times}$   \\
  FPGA 4 CUs  & $\SI{293}{\mega\hertz}$ & $58\%$ & $42\%$  & $\SI{1202}{\mega\op\per\second}$ & $5.3{\times}$ & $190.9{\times}$   \\ \bottomrule
  \end{tabular}}
  \caption{Our 1024-bit (960-bit mantissa) floating-point multiplier executed in hardware, compared to MPFR executed fully in L1 cache on a 36-core CPU node.}
  \label{tab:microbenchmark1024}
\end{table}

We compare an increasing number of compute units instantiated on the FPGA against the full 36-core node running MPFR in \tableref{microbenchmark512} and \tableref{microbenchmark1024} for 512 bits (448-bit mantissa) and 1024 bits (960-bit mantissa) of precision, respectively. The 512-bit multiplier fits up 4 times on each SLR for a total of 16 compute units, yielding $\SI{4.8}{\giga\op\per\second}$ for a speedup over the full 36-core Xeon node of $9.8{\times}$, corresponding to a throughput of more than $351{\times}$ CPU cores at $75\%$ CLB usage and $56\%$ DSP usage. The 1024-bit multiplier can be instantiated once per SLR, yielding $\SI{1.2}{\giga\op\per\second}$ for a $5.3{\times}$ speedup over the Xeon node (corresponding to $191{\times}$ CPU cores).

In the following, we will extend our accelerator to perform matrix multiplication, where we can saturate the computational pipeline without artificially removing the memory bound.

%-----------------------------------------------------------------------------------------------------------------------
\subsection{Benchmarking Matrix Multiplication}
%-----------------------------------------------------------------------------------------------------------------------

\noindent We evaluate the accelerator described in \secref{gemm}, where we maximize the number of compute units that can be instantiated within the resource constraints and allowed by routing according to the SLR/DDR bank assignment scheme in \figureref{compute_unit_mapping}. For the CPU comparison, we run the \texttt{El::Gemm} implementation from Elemental, which is parallelized using MPI\@. We use a tile size of $32{\times}32$ for the FPGA compute units, which balances the trade-off between avoiding useless work on sizes that are not a multiple of the tile size with the reduction in required memory bandwidth at larger tile sizes.

\begin{figure}
  \includegraphics[width=\columnwidth]{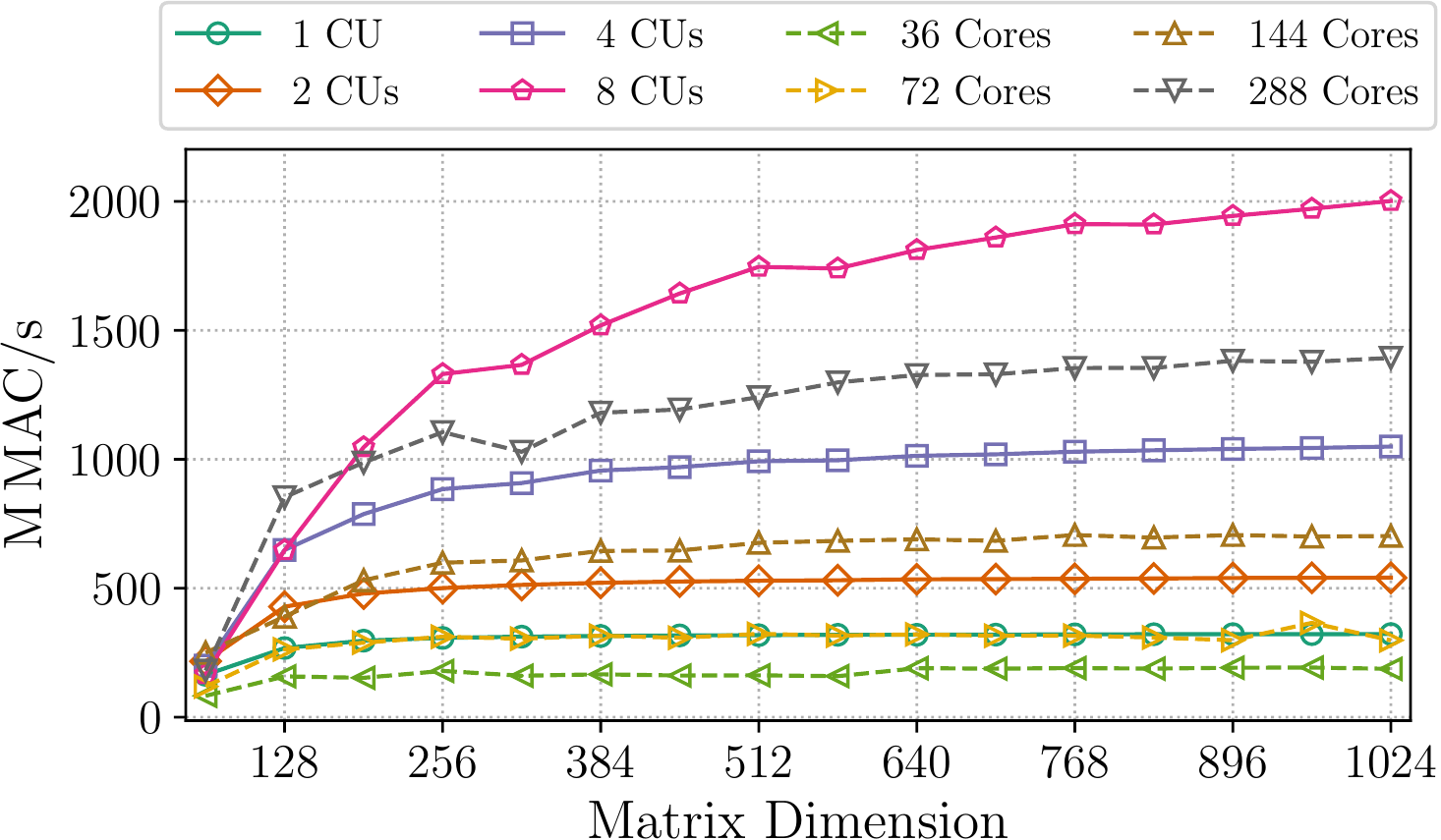}
  \caption{Multiply-addition performance multiplying two matrices of size $n\times n$ with $448$-bit mantissas ($512$ bits total).}
  \label{fig:gemm512}
\end{figure}

\figureref{gemm512} plots the performance of our accelerator for $512$-bit APFP numbers with $448$-bit mantissas against the matrix dimension for $n{\times}n$ matrices for different numbers of replications of the compute unit instantiated on the chip, compared to 1--8 Xeon compute nodes running Elemental/MPFR (dashed lines), in multiply-additions per second (we annotate the more commonly used ``multiply-accumulate'' throughput ($\si{\mega\mac\per\second}$), but note that our addition is not restricted to accumulation). Resource usage is dominated by multiplication, making it the primary constraint on how far we can scale the design (in contrast to machine word-sized floating-point, where additions and multiplications are typically weighted the same when reporting performance). For the MPFR/Elemental performance, we run both $448$-bit and $512$-bit mantissas and take the maximum performance between each pair, to account for performance effects that can occur when the mantissa size is not a power of two.

A single replication of the $512$-bit accelerator exhibits performance corresponding to ${\sim}1{-}2$ Xeon nodes ($60{\sim}$ cores), while the 8-way replicated accelerator corresponds to the throughput of ${>}10{\times}$ Xeon nodes ($375{\times}$ CPU cores). The FPGA GEMM can thus \emph{outperform a small cluster} of dual-socket CPUs, and offers considerable speedup even at small matrix sizes. Introducing more compute units to a fixed size problem (strong scaling along a vertical line in \figureref{gemm512}) reduces the amount of work \emph{per compute unit}, resulting in more replications requiring larger matrix inputs to reach peak performance.
% \johannes{For higher replication counts, the non-monotonic performance difference with increasing matrix size is caused by sizes that are not a multiple of the tile size when partitioned across compute units (e.g., for 4 compute units operating on a matrix with $448$ rows, each compute unit computes $112$ rows, which requires $4$ tiles of size $32$ for a total of $128$ rows, resulting in $15\%$ extra work). The tile sizes can be tuned for typically encountered matrix sizes in the target application to avoid this overhead while not becoming memory bound.}{We are fixing this, should be gone for the final version.}
%
An overview of all designs evaluated is shown in \tableref{designs}, including their logic utilization and the highest performance achieved across different matrix sizes.
Although there is still some resource headroom, further replication is prevented by the number of DDR4 memory interfaces available on the shell used.

\begin{table}
  \resizebox{\columnwidth}{!}{%
    \begin{tabular}{c c | r r r | r}
      \textbf{Precision} & \textbf{CUs} & \textbf{Frequency} & \textbf{CLBs} & \textbf{DSPs} & \textbf{Max. Performance} \\
      \toprule
      $512$ ($448$) & $1$ & $\SI{327}{\mega\hertz}$ & $18.9\%$ & $4.5\%$ & $\SI{322}{\mega\mac\per\second}$ \\
      $512$ ($448$) & $2$ & $\SI{278}{\mega\hertz}$ & $31.7\%$ & $9.0\%$ & $\SI{540}{\mega\mac\per\second}$ \\
      $512$ ($448$) & $4$ & $\SI{278}{\mega\hertz}$ & $46.6\%$ & $14.4\%$ & $\SI{1049}{\mega\mac\per\second}$ \\
      $512$ ($448$) & $8$ & $\SI{293}{\mega\hertz}$ & $65.8\%$ & $35.8\%$ & $\SI{2002}{\mega\mac\per\second}$ \\
      % \midrule
      % $1024$ ($960$) & $1$ & $\SI{131}{\mega\hertz}$ & $35.0\%$ & $7.0\%$ & $\SI{106}{\mega\mac\per\second}$ \\
      \bottomrule
    \end{tabular}%
  }
  \caption{Overview of 512-bit GEMM designs.}
  \label{tab:designs}
\end{table}

%-----------------------------------------------------------------------------------------------------------------------
\subsection{Extending Matrix Multiplication to 1024 bits}
%-----------------------------------------------------------------------------------------------------------------------

\begin{figure}
  \includegraphics[width=\columnwidth]{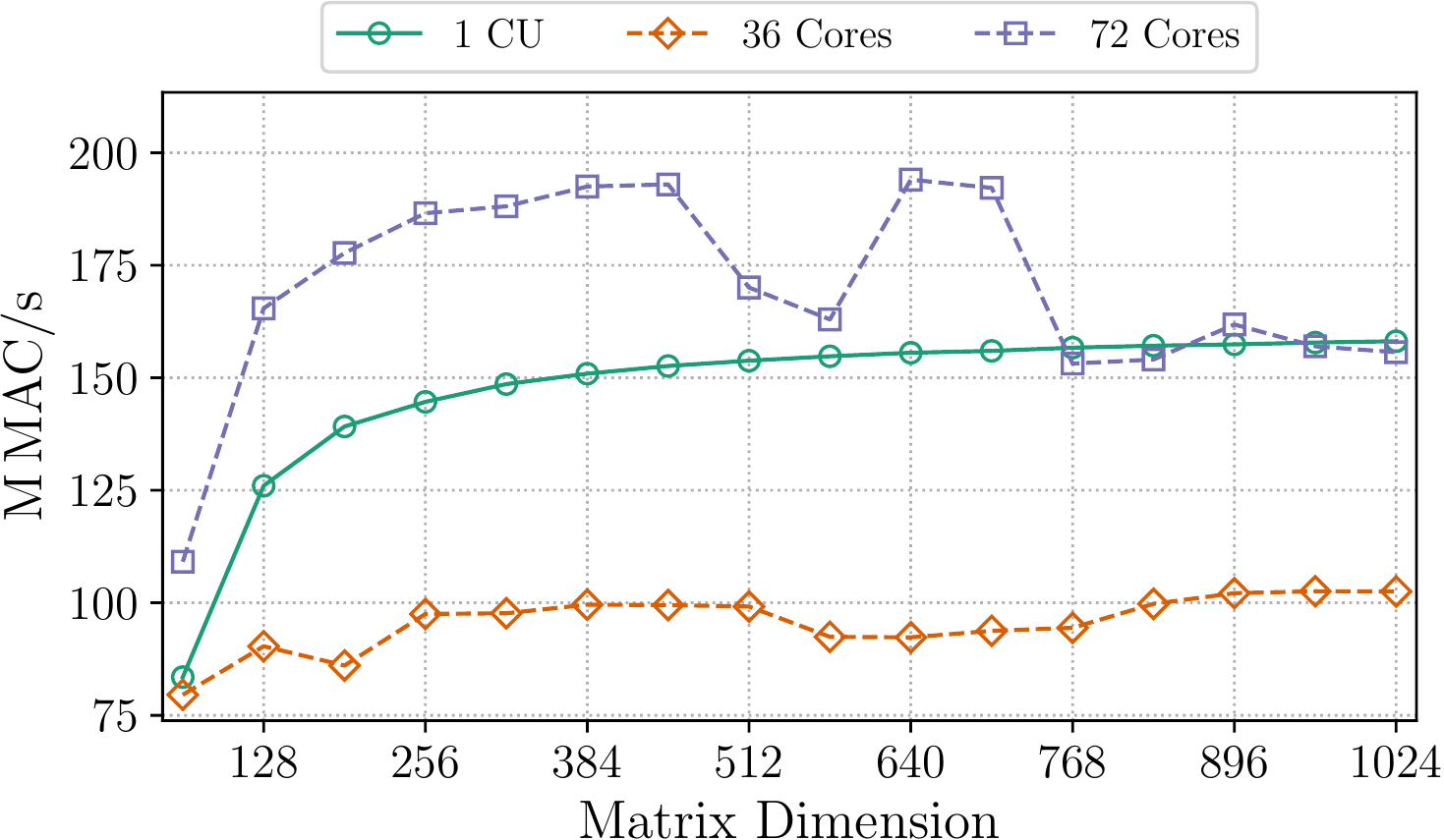}
  \caption{Multiply-addition performance multiplying two matrices of size $n\times n$ with $960$-bit mantissas ($1024$ bits total)}
  \label{fig:gemm1024}
\end{figure}

Extending the matrix multiplier to 1024 bit APFP numbers introduces additional challenges on the target FPGA platform, as a single 1024-bit matrix multiplication compute unit occupies nearly a full SLR on the U250 chip. Based on the results for 512-bit multiplication, two or three 1024-bit multipliers should fit on the device, as this roughly corresponds to six or nine 512-bit multipliers (since each level of Karatsuba decomposition requires 3 half-width multipliers), respectively. However, since these subcomponents are no longer independent and are scheduled as a single pipeline, they are scheduled in a monolithic manner.

We include a preliminary result for 1024-bit (960-bit mantissa) matrix multiplication in \figureref{gemm1024} for a single compute unit. Due to excessive congestion within the multiplication pipeline, the design is downclocked to $\SI{212}{\mega\hertz}$. The throughput at this frequency exceeds the performance of Elemental executed on a 36-core Xeon node, with a peak throughput of $\SI{158}{\mega\mac\per\second}$. At $29.8\%$ CLB utilization, we expect that a more appropriately floorplanned design would allow instantiating $4$ compute units.

% In order to achieve a reasonable floorplan, subcomponents of the 1024-bit multiplication (e.g., individual 512-bit sub-multiplications) should be fixed to distinct SLRs, and connections between them should be sparingly routed across SLRs with ample routing registers to minimize the delay. As of writing, assigning individual functions (or Vitis~HLS ``dataflow functions'') called within a kernel of a computational kernel is not possible in the Vitis toolchain, as only full kernels at the OpenCL-granularity can be fixed to SLRs. A way to mitigate (albeit not circumvent) the SLR assignment issue is to express the HLS kernel as processing elements at a smaller granularity~\cite{hls_transformations}, but such as decomposition is a poor match to the divide-and-conquer approach of Karatsuba multiplication, as every step needs both a prologue and an epilogue before and after the recursion.

% To demonstrate the benefit of FPGA acceleration of arbitrary precision floating-point computations in a real-world applications, we integrate our accelerator with SDPB~\cite{sdpb}, a semidefinite program solver for the conformal bootstrap.

%%%%%%%%%%%%%%%%%%%%%%%%%%%%%%%%%%%%%%%%%%%%%%%%%%%%%%%%%%%%%%%%%%%%%%%%%%%%%%%%%%%%%%%%%%%%%%%%%%%%%%%%%%%%%%%%%%%%%%%%
\section{Related Work}
%%%%%%%%%%%%%%%%%%%%%%%%%%%%%%%%%%%%%%%%%%%%%%%%%%%%%%%%%%%%%%%%%%%%%%%%%%%%%%%%%%%%%%%%%%%%%%%%%%%%%%%%%%%%%%%%%%%%%%%%

% FPGA and GPU acceleration of high precision \chris{workloads has rarely extended beyond double-double (~100 bits) and quad-double (~200 bits) precision \cite{sdpa_campary}.}{There's this weird thing where there's almost nothing > quad-double but a _bunch_ of double-double and quad-double stuff. I don't think we can cite them all?}

Various previous work has proposed accelerators for APFP arithmetics.
CAMPARY~\cite{campary_multi_precision} accelerates up to 424 bits of mantissa using CUDA\@. The authors show up to $19{\times}$ speedup on a Fermi-based Tesla C2075 GPU over a consumer-grade quad-core Sandy Bridge CPU running MPFR, dropping to ${\sim}1{\times}$ for 424-bit mantissas. % Extrapolating this result to a server-grade CPU-node of a similar architectural generation would close the gap at this precision.
MPRES-BLAS~\cite{mpres_blas} presents GPU acceleration of APFP dense linear algebra, showing ${\sim}2{\times}$ speedup over CAMPARY for GEMM, reporting ${\sim}100$-$\SI{120}{\mega\op\per\second}$ for 424-bit precision on a GTX~1080 GPU\@.
Lei~et~al.~\cite{vliw_apfp} implement an APFP accelerator on a Virtex~6 FPGA and report $11.6{\times}$ speedup for $1024$-bit multiplication over MPFR running on a dual-core Core~i3 530 Clarkdale-based CPU\@. % This gap would be significantly reduced at the core count of a server-grade CPU, and the evaluated CPU does not expose Intel~ADX and BMI2 instructions to accelerate arbitrary precision computations.
Lu~et~al.~\cite{gpu_arbitrary_precision} accelerate 500-2000 digits of precision on a GTX~280 GPU on the Tesla architecture and compare it to a quad-core Kentfield CPU running ARPREC~\cite{arprec}, reporting $8{-}9{\times}$ speedup on multiplication. As of writing, the source code published by the authors has not been updated to support modern GPUs.
Common for the above work is that comparisons are made to consumer-grade CPUs, which lack the core count of the server-grade CPUs that are typically employed for larger-scale numerical workloads. Furthermore, Broadwell-based CPUs and onwards received support for the Intel~ADX instruction set in addition to BMI2 introduced with Haswell, which significantly increases CPU performance on arbitrary precision workloads.
Chow~et~al.~\cite{karatsuba_fpga} implement a Montgomery multiplier for modular arithmetic based on Karatsuba decomposition. The authors estimate that $\SI{400}{\mega\op\per\second}$ of Montgomery multiplication throughput is achievable on a Virtex-6 FPGA based on synthesis results, but do not build their design for execution in hardware.

Based on the results presented in this work, our FPGA-based accelerator outperforms all the above accelerators in terms of absolute throughput in hardware, and in terms of speedup when executed in hardware relative to server-grade CPUs of each corresponding generation of hardware at the time of their publication. Furthermore, our work is published as a configurable HLS-based implementation, which can dynamically scale performance by replicating compute units, and compiles for any Vitis/XRT-based Xilinx platform.

% Applications such as SDPB necessitate accelerating precisions of several hundreds to a few thousand bits. In recent years, there have been a handful of multiprecision BLAS-like software packages such as \cite{mpres_blas, elemental, mlapack, mpmath} of which only \cite{mpres_blas} utilizes an accelerator (GPU).

% \chris{Julia?}{Julia kinda transparently handles multiprecision so most applications written in it just *do* multiprecision.}

%%%%%%%%%%%%%%%%%%%%%%%%%%%%%%%%%%%%%%%%%%%%%%%%%%%%%%%%%%%%%%%%%%%%%%%%%%%%%%%%%%%%%%%%%%%%%%%%%%%%%%%%%%%%%%%%%%%%%%%%
\section{Conclusion}
%%%%%%%%%%%%%%%%%%%%%%%%%%%%%%%%%%%%%%%%%%%%%%%%%%%%%%%%%%%%%%%%%%%%%%%%%%%%%%%%%%%%%%%%%%%%%%%%%%%%%%%%%%%%%%%%%%%%%%%%

In this work, we showed how FPGAs provide an excellent platform for accelerating fundamental operators for arbitrary precision floating point (APFP) arithmetic.
We present a deeply pipelined design implementing APFP multiplication using a Karatsuba decomposition bottoming out at naive multiplication in DSPs, yielding a multiplication throughput of up to $\SI{4.8}{\giga\op\per\second}$ for 512-bit and $\SI{1.2}{\giga\op\per\second}$ for 1024-bit numbers on an Alveo U250 accelerator, corresponding to the throughput of more than $351{\times}$ and $191{\times}$ CPU cores running MPFR, respectively. We combine the multiplier with our APFP adder to perform general matrix-matrix multiplication in hardware, showing $\SI{2.0}{\giga\mac\per\second}$ on 512-bit numbers, which corresponds to the throughput of more than $375{\times}$ CPU cores, matching the performance of a 10-node Xeon cluster.
For numerical codes that are dominated by arbitrary precision arithmetic, such as semidefinite program (SDP) solvers, we expect these gains to translate into real-world speedups on applications such as the conformal bootstrap studying phase transitions in quantum field theory.
To facilitate this, we publish the accelerator code as an open-source HLS-based project, configurable for any Vitis/XRT-supported Xilinx FPGA\@. We provide a plug-and-play software interface that can be dropped into existing numerical codes, allowing scientists to tap into FPGA acceleration of APFP with minimal code changes.

%%%%%%%%%%%%%%%%%%%%%%%%%%%%%%%%%%%%%%%%%%%%%%%%%%%%%%%%%%%%%%%%%%%%%%%%%%%%%%%%%%%%%%%%%%%%%%%%%%%%%%%%%%%%%%%%%%%%%%%%
\section*{Acknowledgments}
%%%%%%%%%%%%%%%%%%%%%%%%%%%%%%%%%%%%%%%%%%%%%%%%%%%%%%%%%%%%%%%%%%%%%%%%%%%%%%%%%%%%%%%%%%%%%%%%%%%%%%%%%%%%%%%%%%%%%%%%

\setlength{\intextsep}{1pt}%
\setlength{\columnsep}{5pt}%
\begin{wrapfigure}{l}{.13\columnwidth}
  \includegraphics[width=.13\columnwidth]{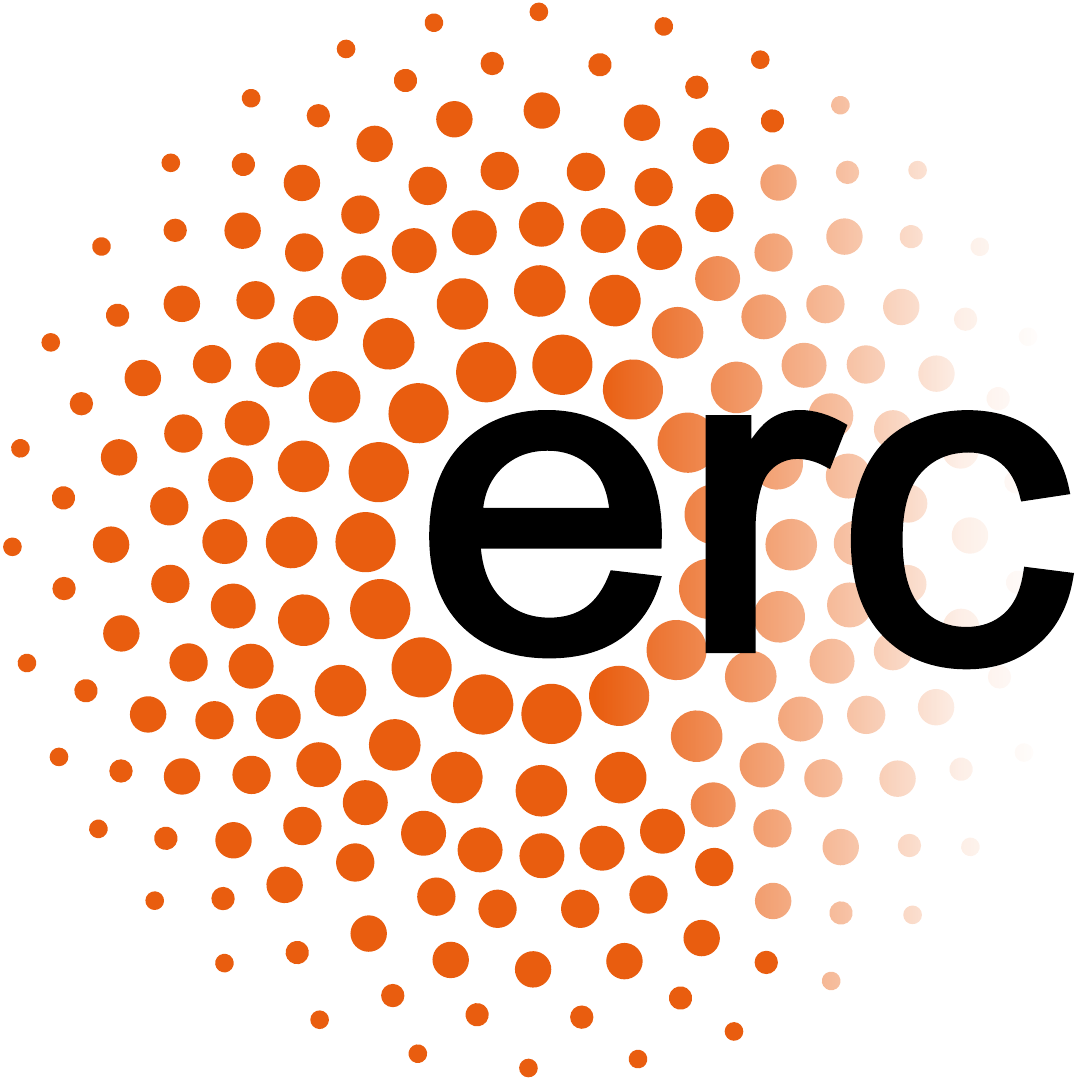}
\end{wrapfigure}
\noindent This project has received funding from the European Research Council (ERC) under the European Union's Horizon 2020 research and innovation programme grant agreeement no.\ 101002047 and from the European High-Performance Computing Joint Undertaking (JU) under grant agreement no.\ 101034126. Christopher A.  Pattison is supported by Air Force Office of Scientific Research (AFOSR), FA9550-19-1-0360, and thanks Dustin Kenefake for inspiring discussions. David Simmons-Duffin is supported by Simons Foundation grant 488657 (Simons Collaboration on the Nonperturbative Bootstrap) and a DOE Early Career Award under grant no.\ DE-SC0019085.

% Bibliography
\bibliographystyle{IEEEtran}
\bibliography{apfp}

\end{document}